\documentclass[a4paper,12pt]{article}
\pdfoutput=1
\usepackage{jheppub}
\usepackage{amsmath}
\usepackage{amsxtra}
\usepackage{amssymb}
\usepackage{graphicx}
\usepackage{pifont}
\usepackage{dcolumn}
\usepackage{bm}
\usepackage{multirow}
\usepackage[font=scriptsize,labelfont=bf]{caption}

\newcommand{\stau}{\tilde{\tau}}

\newcommand{\neut}{\tilde{\chi}^0_1}
\newcommand{\nneut}{\tilde{\chi}^0_2}
\newcommand{\nnneut}{\tilde{\chi}^0_3}
\newcommand{\nnnneut}{\tilde{\chi}^0_4}

\newcommand{\charg}{\tilde{\chi}^+_1}
\newcommand{\ccharg}{\tilde{\chi}^+_2}

\begin{document}

\title{Non-universal gaugino mass GUT models in the light of dark matter and LHC constraints}
\author[a]{Joydeep Chakrabortty,}
\affiliation[a]{Department of Physics, Indian Institute of Technology, Kanpur-208016, India}
\author[b]{Subhendra Mohanty,}
\affiliation[b]{Physical Research Laboratory, Ahmedabad-380009, India}
\author[c]{Soumya Rao}
\affiliation[c]{Department of Theoretical Physics and Centre for
Theoretical Sciences, Indian Association for the Cultivation of Science,\\
2A \& 2B Raja S.C. Mullick Road, Kolkata-700032, India.}
\emailAdd{joydeep@iitk.ac.in,mohanty@prl.res.in,tpsr2@iacs.res.in}

\abstract{ We perform a comprehensive study of  $SU(5)$, $SO(10)$ and $E(6)$
	supersymmetric GUT models where the gaugino masses are generated through the
	F-term breaking vacuum expectation values of the non-singlet scalar fields. In
	these models the gauginos are non-universal at the GUT scale unlike in the mSUGRA
	scenario. We discuss the properties of the LSP which is stable and a viable
	candidate for cold dark matter.  We look for the GUT scale parameter space that
	leads to the the lightest SM like Higgs mass in the range of 122-127 GeV
	compatible with the observations at ATLAS and CMS , the relic density in the
	allowed range of WMAP-PLANCK and compatible with other constraints from colliders
	and direct detection experiments. We scan universal scalar ($m_0^G$), trilinear
	coupling $A_0$ and  $SU(3)_C$ gaugino mass ($M_3^G$) as the independent  free
	parameters  for these models. Based on the gaugino mass ratios at the GUT scale,
	we classify  25 SUSY GUT models and find that of these only 13 models satisfy the
	dark matter and collider constraints.  Out of these 13 models there is only one
	model where there is a  sizeable SUSY contribution to muon $(g-2)$.
}

\maketitle
\flushbottom

\section{Introduction}

Supersymmetry (SUSY) is an aesthetically  appealing model which  provides a natural
mechanism to stabilise the Higgs mass and solves the gauge hierarchy problem of the
Standard Model. The general Supersymmetric Standard Model at the electroweak scale has
more than a hundred  parameters which make the  predictability  of such models
questionable.  An economical Supersymmetric Standard Model can be constructed  which
contains only  a few  free parameters known as the constrained Minimal Supersymmetric
Standard Model (cMSSM), which relates to the high scale minimal supergravity models
(mSUGRA) through renormalisation groups.  In mSUGRA there are only 5 parameters: universal
scalar mass $m_0$, universal gaugino mass $M_{1/2}$, $\tan \beta$, sign of $\mu$
($sgn(\mu)$) and universal tri-linear couplings $A_0$. The lightest supersymmetric
particle (LSP) is mostly bino-like.  But the recent LHC data is ruling out most of its
parameter space for obtaining the WMAP-PLANCK measured relic density of bino as cold dark
matter. But it is not necessary to have all the gauginos unified at the unification scale.

It has been noted in \cite{Ellis:1985jn, Drees:1985bx,Anderson:1996bg,
Anderson:1999uia,Huitu:1999vx, Corsetti:2000yq, Chattopadhyay:2003yk} that in
Supersymmetric Grand Unified Theories (SUSY GUTs) the boundary conditions at the high
scale itself can be different than that in mSUGRA. The gaugino masses can be non-universal
at the GUT scale itself. The renormalisation group evolutions (RGEs) further change their
ratios at the electroweak scale and thus the phenomenology of such models can be
completely different compared to mSUGRA.  But these non-universalities in SUSY GUT models
are completely determined from the group theoretic structure of the symmetry breaking
scalar fields. In \cite{Ellis:1985jn, Drees:1985bx, Anderson:1996bg, Anderson:1999uia,
Huitu:1999vx, Corsetti:2000yq, Chattopadhyay:2003yk} these non-universal gaugino mass
ratios were first calculated for $SU(5)$ group with 24-, 75-, and 200- dimensional scalar
fields. Later in \cite{cr1, martin1, subho-jd,cr2} the non-universal gaugino mass ratios
are presented for all possible breaking patterns having all possible scalar fields for
$SO(10)$, and $E(6)$ GUT gauge groups.

These non-universal models are clear departure from mSUGRA in the boundary conditions.
Thus different non-universalities lead to different kind of LSP scenarios. Some recent
papers  has partly grabbed the impact of non-universality feature in either minimal or
non-minimal version of models in the context of dark matter search, see for example
\cite{Cerdeno:2004zj, Huitu:2005wh, King:2007vh, Huitu:2008sa, debottam, Feldman:2009zc,
Gogoladze:2009bn, Holmes:2009uu, subho-jd, Okada:2011wd, Bhattacharyya:2011se,
Amsel:2011at, Gogoladze:2011ug, Younkin:2012ui, Baer:2012by, Caron:2012sf, Miller:2013jra,
Badziak:2013eda, Han:2013gba, Bhattacharya:2013uea}.

In this paper we have encapsulated the parameter space for all models (25)  arising from
different GUT gauge groups, like $SU(5)$, $SO(10)$, and $E(6)$ and the symmetry breaking
patterns from all the possible scalar representations which can break the F-term and gauge
symmetries as well. These give rise to different mass ratios of the three gauginos at the
GUT scale. Here we have considered only those models for which all of them are non-zero at
the unification scale.

Running the masses down to the electroweak scale we get the ratios $M_1:M_2:M_3$ for
different models which are quite distinct from the mSUGRA relation $1:2:6.7$ at
electroweak scale. Here $M_1,M_2,M_3$\footnote{We define the GUT scale input of these
parameters as $M_i^G$.} are the gaugino masses corresponding to the $U(1)_Y, SU(2)_L,
SU(3)_C$ gauge groups respectively. We scan the parameters $M_3^G, m_0^G, A_0, \tan \beta$
and test the range of parameters for each model which give the lightest Higgs mass in the
range $122 ~\mbox{GeV} < M_h < 127 ~\mbox{GeV}$ \cite{higgs-125}, and the dark matter
relic density within 3-sigma of the WMAP-PLANCK \cite{wmap,planck} measured band
$0.112<\Omega h^2 <0.128$.  In addition we have other constraints: within the allowed
parameter space the contribution to the $B_s \to X_s \gamma$ \cite{b2sg}, $B_s \to \mu^+
\mu^-$ \cite{bs2mupm} and the muon $(g-2)$ \cite{mug-2} must satisfy the experimental
bounds.  We have also set the lower limit on the gluino mass ($m_{\tilde{g}}$) to be
$1.4$ TeV\footnote{Though it is not playing any crucial role in our analysis as within the
parameter space allowed by the other constraints $m_{\tilde{g}}$ is more than 1.8 TeV or
so.}.  Once these criterion are satisfied we compute the best fit value for the SUSY
contribution to muon $(g-2)$ within the parameter space of the models constrained by the
other experimental limits.

Of the 25 models examined we find that only 13 models satisfy the collider and dark matter
experimental constraints and we find however that none of these 13 models explain the
experimental value of muon $(g-2)$ \cite{mug-2}. The other 12 models are mainly ruled
out when we impose light Higgs mass and 3-sigma relic density constraints together. The
largest contribution to muon $(g-2)$ comes from the the models where the gaugino mass
ratio at GUT scale is $M_1:M_2:M_3 \equiv -1/2:-3/2:1$ and this model has a bino like dark
matter with mass 177 GeV.

There are five wino, five bino and three higssino dark matter  models which give the
WMAP-PLANCK relic density.  Some of the models can be probed by the  XENON1T \cite{Xeno1T}
and Super-CDMS \cite{superCDMS} experiments and one model is ruled out by XENON100
\cite{xenon100}.

\section{SUSY GUT and non-universal gauginos }

Supersymmetry and Grand Unified Theory  both have different motivations to be suitable
theories beyond the Standard Model. Supersymmetry justifies the gauge hierarchy problem
and predicts many other superpartners of SM particles. In R-parity conserving SUSY
theories LSP is stable and can be a viable cold dark matter candidate. Here we will focus
only on the neutralino LSPs. Within this framework SUSY is expected to explain the
observed relics of the Universe. Added with these nice outcomes the extra feature of this
theory in the GUT framework is very encouraging. SUSY improves the gauge coupling
unification in most of the GUT models.  Thus SUSY GUT models are phenomenologically
interesting and motivating.

The GUT symmetry is broken when a non-singlet direction under that gauge group acquires
vacuum expectation value.  In  SUSY GUT unified frame work most of the couplings (masses)
are degenerate at the unification  scale. In its minimal form all the gauginos and scalars
are universal respectively. The other free parameters are tri-linear coupling ($A_0$)
which is also universal, $\tan \beta$ (ratio of the vacuum expectation values, $vev$, of
two Higgs doublets), and sign of $\mu$ (Higgs parameter). But we can have other
possibilities, like gauginos or scalars are non-universal at the High scale themselves
when we work under SUSY GUT framework. The scalars that cause the GUT symmetry breaking
may develop a F-term breaking $vev$.  Thus GUT and supersymmetry are broken via a single
scalar but through the $vev$s in different directions. The gauge kinetic term can be
recast in a much simpler form as: $\frac{\eta}{M}Tr(F_{\mu \nu} \Phi F^{\mu \nu})$ where
$\eta$ is dimensionless parameter, $M=M_{Pl}/\sqrt{8\pi}$ (reduced Planck mass). As
$F_{\mu \nu}$ transforms as adjoint of the unbroken GUT groups, $\Phi$ belongs to the
symmetric product of the two adjoints.

In this paper we have worked on $SU(5)$, $SO(10)$, $E(6)$ GUT groups, thus the choices of
scalars are as following:
\begin{eqnarray}
 SU(5) & \Rightarrow (24\otimes 24)_{sym} & = 1 \oplus 24 \oplus 75 \oplus 200, \nonumber \\
 SO(10) & \Rightarrow (45\otimes 45)_{sym} & = 1 \oplus 54 \oplus 210 \oplus 770, \\
 E(6) & \Rightarrow (78\otimes 78)_{sym} & = 1 \oplus 650  \oplus 2430, \nonumber
\end{eqnarray}
where $24,45,78$ are the dimensions of the adjoint representations of $SU(5)$, $SO(10)$,
$E(6)$ respectively.

\begin{table}[htb]
\centering
\resizebox{\textwidth}{!}{
\begin{tabular}[c]{|c|c|c|c|}
\hline
Model Number &$M_1$ : $M_2$ : $M_3$ &$M_1$ : $M_2$ : $M_3$ & Model \\
 & (at M$_X$) & (at M$_{EW}$) & \\
\hline\hline
 1 & $-19/5 $ : 1 : 1 & $-19/5 $ : 2 : 6 & $SO(10)\xrightarrow{(1,0)\subset 210}(SU(5)\otimes U(1))_{flipped}$\\
\hline
2 & -3 : 1 : 1 & -3 : 2 : 6 & $E(6)\xrightarrow{(189,1)\subset 2430,650}(SU(6)\otimes SU(2)_X)$\\
\hline
3 &  $-13/5$ : 1 : 1 &  $-13/5$ : 2 : 6 &  $E(6)\xrightarrow{(1,1)\subset 650}(SU(6)\otimes SU(2)_R)$ \\
\hline
4 &  $-22/5$ : 1 : 1  &  $-22/5$ : 2 : 6 &  $E(6)\xrightarrow{(1,0)\subset 650}(SO(10)\otimes U(1))_{flipped}$\\
\hline
5 &  $41/15$ : 1 : 1  &  $41/15$ : 2 : 6  & $E(6)\xrightarrow{(1,1)\subset 2430}(SU(6)\otimes SU(2)_R)$\\
\hline
6 &  122/5 : 1 : 1 &  122/5 : 2 : 6  &  $E(6)\xrightarrow{(1,0)\subset 2430}(SO(10)\otimes U(1))_{flipped}$\\
\hline
7 &  -101/10 : -3/2 : 1 &  -101/10 : -3 : 6 & $SO(10)\xrightarrow{(24,0)\subset 770}(SU(5)\otimes U(1))_{flipped}$  \\
\hline
8 &  77/5 : 1 : 1 &  77/5 : 2 : 6  & $SO(10)\xrightarrow{(1,0)\subset 770}(SU(5)\otimes U(1))_{flipped}$\\
\hline
9 &  10 : 2 : 1  &  10 : 4 : 6 & $SO(10)\xrightarrow{(200)\subset 770} SU(5)$ \\
\hline
\end{tabular}
}
\caption{Ratios of gaugino masses that lead to $M_1>M_2$ at EWSB($M_{EW}$) Scale.}
\label{tabm1}
\end{table}

\begin{table}[htb]
\centering
\resizebox{\textwidth}{!}{
\begin{tabular}[c]{|c|c|c|c|}
\hline
Model Number &$M_1$ : $M_2$ : $M_3$ & $M_1$ : $M_2$ : $M_3$ & Model \\
 & (at M$_X$) & (at M$_{EW}$) & \\
\hline\hline
10 &  $\frac{9}{5}$ : 1 : 1 &  $\frac{9}{5}$ : 2 : 6 &  $E(6)\xrightarrow{(405,1)\subset 2430}(SU(6)\otimes SU(2)_R)$  \\
\hline
11 &  -5 : 3 : 1   &  -5 : 6 : 6 &  $SO(10)\xrightarrow{(75)\subset 770} SU(5)$   \\
\hline
12 &  1 : 35/9 : 1 &  1 : 70/9 : 6 & $E(6)\xrightarrow{(1,1)\subset 2430}(SU(6)\otimes SU(2)_L)$   \\
\hline
13 &  1 : -5 : 1 &  1 : -10 : 6 & $E(6)\xrightarrow{(1,1)\subset 650}(SU(6)\otimes SU(2)_L)$   \\
\hline
14 &  -3/5 : 1 : 1 &  -3/5 : 2 : 6 & $E(6)\xrightarrow{(189,1)\subset 650,2430}(SU(6)\otimes SU(2)_R)$   \\
\hline
15 &  -1/5 : -1 : 1  &  -1/5 : -2 : 6  & $E(6)\xrightarrow{(35,1)\subset 650}(SU(6)\otimes SU(2)_R)$  \\
\hline
16 &  1/10 : 5/2 : 1  &  1/10 : 5 : 6  & $E(6)\xrightarrow{(770,0)\subset 2430}(SO(10)\otimes U(1))_{flipped}$   \\
\hline
17 &  1/10 : -3/2 : 1 &  1/10 : -3 : 6  & $E(6)\xrightarrow{(54,0)\subset 650}(SO(10)\otimes U(1))_{flipped}$   \\
\hline
18 &  2/5 : 2 : 1   &  2/5 : 4 : 6 &  $E(6)\xrightarrow{(770,0)\subset 2430}(SO(10)\otimes U(1))_{flipped}$  \\
 & & & $(SO(10)\otimes U(1))_{flipped}\xrightarrow{(200)\subset 770}SU(5)$ \\
\hline
19 &  -1/5 : 3 : 1  &  -1/5 : 6 : 6  &  $E(6)\xrightarrow{(210,0),(770,0)\subset 650,2430}(SO(10)\otimes U(1))_{flipped}$  \\
 & & & $(SO(10)\otimes U(1))_{flipped}\xrightarrow{(75)\subset 210,770}SU(5)$ \\
\hline
20 &  5/2 : -3/2 : 1  &  5/2 : -3 : 6  &  $E(6)\xrightarrow{(770,0)\subset 2430}(SO(10)\otimes U(1))_{flipped}$  \\
 & & & $(SO(10)\otimes U(1))_{flipped}\xrightarrow{(24)\subset 770}SU(5)$  \\
\hline
21 &  -1/5 : -3/2 : 1 &  -1/5 : -3 : 6  &  $E(6)\xrightarrow{(210,0)\subset 650,2430}(SO(10)\otimes U(1))_{flipped}$  \\
& & & $(SO(10)\otimes U(1))_{flipped}\xrightarrow{(24)\subset 210}SU(5)$ \\
\hline
22 &  -1/5 : 1 : 1  &  -1/5 : 2 : 6 &  $E(6)\xrightarrow{(210,0)\subset 650,2430}(SO(10)\otimes U(1))_{flipped}$  \\
 & & & $(SO(10)\otimes U(1))_{flipped}\xrightarrow{(1)\subset 210}SU(5)$ \\
\hline
23 &  19/10 : 5/2 : 1  &  19/10 : 5 : 6 &  $SO(10)\xrightarrow{(1,1)\subset 770}(SU(4)\otimes SU(2)_R$   \\
\hline
24 &  -1/2 : -3/2 : 1 &  -1/2 : -3 : 6   &  $SO(10)\xrightarrow{(24)\subset 54,210,770} SU(5)$    \\
  &  &  & $SO(10)\xrightarrow{(24,0)\subset 54}(SU(5)\otimes U(1))_{flipped}$\\
  & & &  $SO(10)\xrightarrow{(1,1)\subset 54}(SU(4)\otimes SU(2)_R)$\\
\hline
25 &  7/10 : -3/2 : 1  &  7/10 : -3 : 6    &  $SO(10)\xrightarrow{(24,0)\subset 210} (SU(5)\otimes U(1))_{flipped}$    \\
\hline
\end{tabular}
}
\caption{Ratios of gaugino masses that lead to $M_1<M_2$ at EWSB($M_{EW}$) Scale.}
\label{tabm2}
\end{table}

It has been noted earlier that these operators also change the gauge coupling unification
conditions at the high scale and in many cases it improves the unifications, see for
example \cite{Shafi:1983gz,Hill:1983xh,Calmet:2008df,Chakrabortty:2009xm}.  As these
scalars are non-singlet, their $vev$ treat the SM gauginos in different footing. Thus the
SM gauge fields, i.e. the gauge couplings are scaled differently. These types of operators
can inject non-universality in the gaugino masses.

In $SU(5)$ models with only possible breaking pattern: $SU(5) \to SU(3) \otimes SU(2)
\otimes U(1)$  the  scalar fields of 24, 75 and 200 dimensions lead to three different set
of non-universal gaugino mass ratios. But as the ranks of $SO(10)$ and $E(6)$ are larger
than that of the SM there are more than one possible breaking patterns of these GUT
symmetry groups. We have noted the gaugino mass ratios for the following intermediate
breaking patterns of $SO(10)$: $SU(5)\otimes U(1), SU(4)\otimes SU(2)\otimes SU(2)$, and
for $E(6)$ we have considered $ SO(10)^{'}\otimes U(1), SU(3)\otimes SU(3)\otimes SU(3),
SU(6)\otimes SU(2)$. Though the group theoretic structures are similar in few cases but as
the SM symmetry is realised in different ways the non-universal gaugino mass ratios are
different for those models. For example $SU(5) \otimes U(1)$ is a maximal subgroup of
$SO(10)$. In normal $SU(5)$ model the extra $U(1)$ does not contribute in $U(1)_Y$ of SM,
but in flipped $SU(5)$ model the hypercharge generator of SM is a linear combination of
this $U(1)$ and another Abelian group coming from $SU(5)$.  In these two cases the ratio
of the gaugino masses at the GUT scale are different from each other.  Here we have
tabulated 24 different types of non-universal gaugino mass ratios discarding the
possibility of one of the gauginos has zero mass at the high scale.  It is very
interesting to note that unlike the mSUGRA scenario here we can have either $M_1 > M_2$ or
$M_1 < M_2$ and even $M_1 \simeq M_2$ at the electroweak scale. Thus where in mSUGRA we
have mostly bino-like Lightest Supersymmetric Particle (LSP), in these SUSY-GUT frame work
because of the non-universality one can have purely bino-  or  wino-  or  higgsino-
dominated LSP or a mixed one also.

Here we briefly mention our model identifications depending on the GUT groups, choices of
scalar fields and symmetry breaking patterns, see Tables~\ref{tabm1} and \ref{tabm2}.
Here we would like to pass a remark that while calculating these gaugino mass ratios for
different models it has been assumed that all the intermediate symmetry scales are same as
the unification (GUT) scale, i.e., the GUT symmetry is broken to the SM gauge group at the
unification scale  itself.

\section{Results}

We examine the different non-universal gaugino mass models in the light of relic density,
direct detections and collider bounds. We have classified all the models in three
categories depending on the compositions of the LSPs: bino-dominated, wino-dominated, and
higgsino-dominated.

\subsection{Relic density and collider constraints}\label{RD_LEC}
We have used the following constraints in our analysis and determine which of the 25
models arising from non-singlet Higgs pass these tests:
\begin{enumerate}
	\item Higgs mass bound from LHC \cite{higgs-125}
		\[122\mbox{ GeV} < M_h < 127\mbox{ GeV}\]
   \item Relic density constraint from WMAP-PLANCK data at $3\sigma$ \cite{wmap,planck}
		\[0.1118 < \Omega\mbox{h}^2 < 0.1280\]		

   \item Gluino mass ($m_{\tilde{g}}$) $>$ 1.4 TeV.
   		
	\item Branching fraction for $B_s\to X_s\gamma$ at $2\sigma$ \cite{b2sg}
		\[3.05\times 10^{-4} < \mbox{BR}(B_s\to X_s\gamma) < 4.05\times 10^{-4}\]
		
	\item Branching fraction for $B_s\to\mu^+\mu^-$ at $2\sigma$ \cite{bs2mumu}
		\[0.8\times 10^{-4} < \mbox{BR}(B_s\to\mu^+\mu^-) < 6.2\times 10^{-4}\]

	\item Ratio of branching fraction for $B_u\to\tau\nu_\tau$ in MSSM to that in SM
		at $3\sigma$ \cite{b2taunu}
		\[0.46 < \frac{\mathrm{BR}(B_u\to\tau\nu_\tau)_{MSSM}}{\mathrm{BR}(B_u\to\tau\nu_\tau)_{SM}} < 1.78\]

	\item  There is a discrepancy in anomalous muon magnetic moment, $a_\mu\equiv (g-2)/2$,
        between experimental value \cite{mug-2} and SM prediction
        \cite{g-2},
        \[\Delta a_\mu= a_{\mu}^{exp}-a_{\mu}^{SM} =( 26.1 \pm 8.0)\times 10^{-10} \]
 We compute the SUSY contribution  to $ a_\mu$ for each of the models which satisfies the other criterion
listed above. We find only one model where there is a substantial SUSY contribution with $
a_{\mu}^{SUSY}=2.65 \times 10^{-10}$.

\end{enumerate}
For our analysis we use the two-loop RGE code {\tt SuSpect} \cite{suspect} to obtain the
weak scale SUSY particle spectrum.  In addition we use the MicrOMEGAs code
\cite{micromega} to evaluate low energy constraints like $B_s\to\mu^+\mu^-$, $B_s\to
X_s\gamma$, muon $(g-2)$ and relic density.  The parameter scan performed in this analysis
takes the following ranges of parameters :
\begin{align*}
	m_0\in[100,2000]&\mbox{ GeV},\\
	M_3^G\in[800,2000]&\mbox{ GeV},\\
sgn(\mu) \equiv +,- &.
	\label{prange}
\end{align*}
Here we define $M_3$ as $M_3^G$ at GUT scale and other gaugino masses $M_1,M_2$ are set by
the gaugino mass ratios at that scale.  We have performed our analysis for three
different choices of tri-linear coupling $A_0=-1,0,1$ TeV.  We have chosen $\tan\beta=10$
unless mentioned otherwise.

\begin{table}[htp]\label{bp_GUT}
	\centering
	\resizebox{\textwidth}{!}{
	\begin{tabular}{|c|c|c|c|c|c|c|}
                \hline\hline
		Model no. & $M_1^G:M_2^G:M_3^G$ & $m_0$ (GeV) & $M_3^G$ (GeV) & $A_0$
		(TeV) & $\tan\beta$ & sgn$(\mu)$ \\ \hline
		1 & $-\frac{19}{5}:1:1$ & $182$ & $2038$ & $-1$ & $10$ & $+$ \\ \hline
		2 & $-3:1:1$ & $100$ & $1620$ & $-1$ & $10$ & $+$ \\ \hline
		3 & $-\frac{13}{5}:1:1$ & $300$ & $1320$ & $-1$ & $10$ & $+$ \\ \hline
		4 & $-\frac{22}{5}:1:1$ & $130$ & $2055$ & $-1$ & $10$ & $+$ \\ \hline
		5 & $\frac{41}{15}:1:1$ & $300$ & $1460$ & $-1$ & $10$ & $+$ \\ \hline
		9 & $10:2:1$ & $116$ & $966$ & $-1$ & $10$ & $+$ \\ \hline
		10 & $\frac{9}{5}:1:1$ & $1000$ & $1190$ & $-1$ & $10$ & $+$ \\ \hline
		11 & $-\frac{1}{5}:3:1$ & $2000$ & $1650$ & $-4$ & $40$ & $+$ \\ \hline
		18 & $\frac{2}{5}:2:1$ & $200$ & $1119$ & $-1$ & $10$ & $+$ \\ \hline
		19 & $-5:3:1$ & $789$ & $1719$ & $-3.5$ & $10$ & $+$ \\ \hline
		20 & $\frac{5}{2}:-\frac{3}{2}:1$ & $1900$ & $1740$ & $-1$ & $10$ & $-$ \\ \hline
		22 & $-\frac{1}{5}:1:1$ & $150$ & $1355$ & $-1$ & $10$ & $-$ \\ \hline
		24 & $-\frac{1}{2}:-\frac{3}{2}:1$ & $506$ & $800$ & $-3.5$ & $20$ & $-$ \\
		\hline\hline
	\end{tabular}
}
	\caption{Input parameters at GUT scale for the benchmark point chosen for each of
	the $13$ models.  We choose the parameters such that in each case we get a maximal
	contribution from SUSY to muon $(g-2)$.}
	\label{tab:inpar}
\end{table}




\begin{figure}[htp]
\centering
\includegraphics[width=0.49\textwidth]{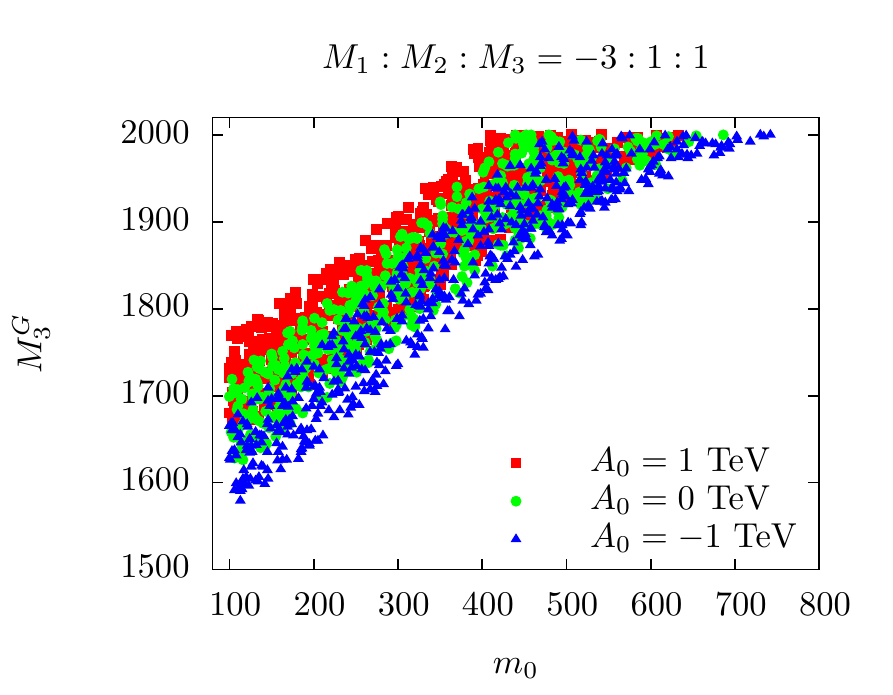}
\includegraphics[width=0.49\textwidth]{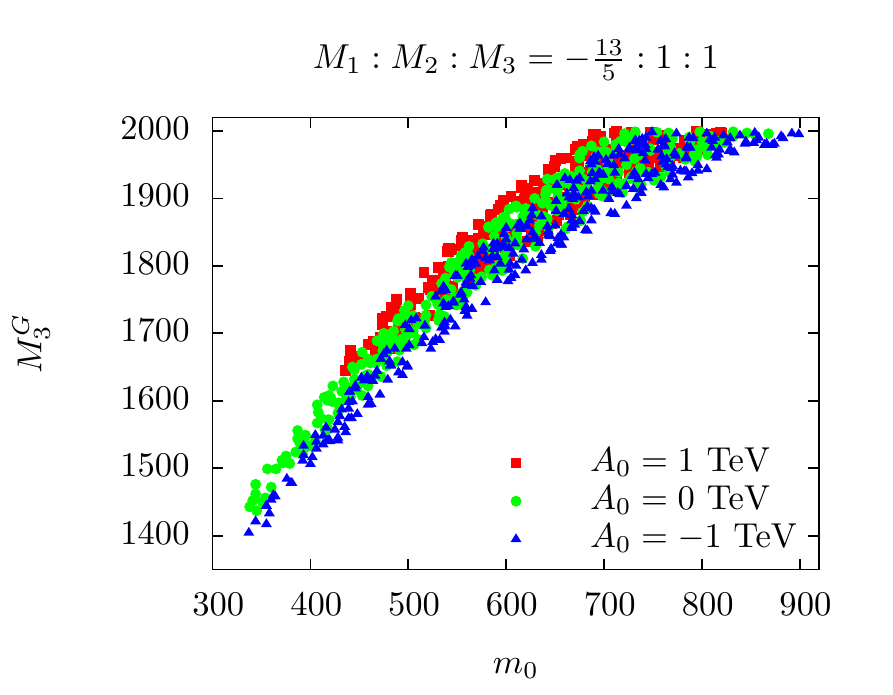}\vskip 0.2cm
\includegraphics[width=0.49\textwidth]{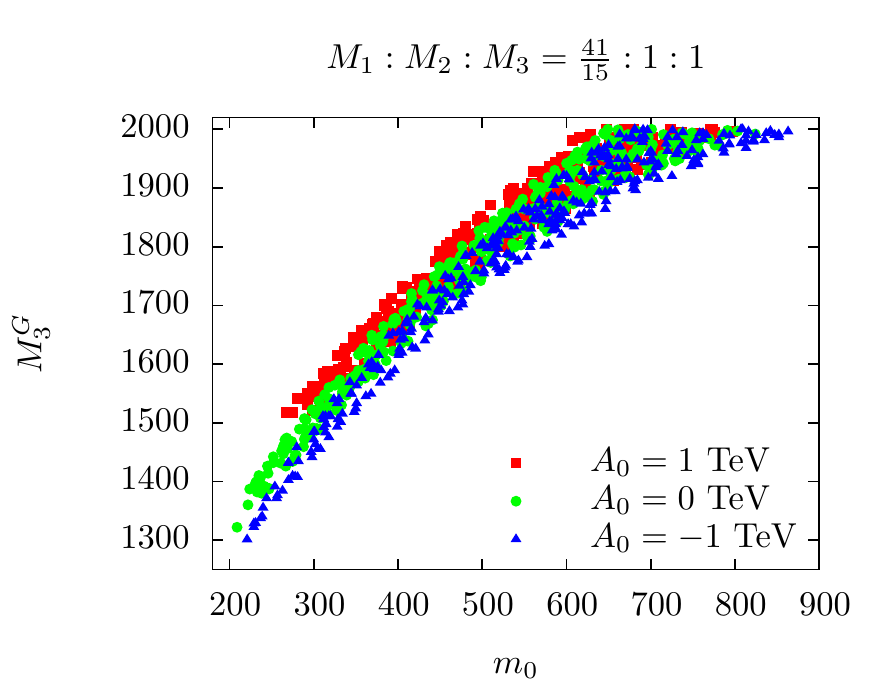}
\caption{The allowed parameter space satisfying all the low energy constraints as listed
	in the text except muon $(g-2)$ for heavy wino DM models with the GUT scale the
	gaugino mass ratios as mentioned on top of each panel.  The choice of other
	parameters are $\tan\beta=10, sgn(\mu)\equiv +ve~(positive)$.   For model
	2$(-3:1:1)$ the allowed mass range for $m_0$ is $\sim 100-700$ GeV for $A_0=0,1$
	TeV  with $M_3^G$ ranging from $\sim 1600-2000$ GeV, whereas for $A_0=-1$ TeV,
	$m_0$ ranges between $\sim 100-750$ GeV with $M_3^G$ between $\sim 1550-2000$ GeV.
	For model 3$(-13/5:1:1)$ the allowed mass range for $m_0$ is $\sim 300-900$ GeV
	with $M_3^G$ between $\sim 1400-2000$ GeV for $A_0=0,-1$ TeV, but for $A_0=1$ TeV,
	$m_0$ ranges between $\sim 400-800$ GeV with $M_3^G$ between $\sim 1600-2000$
	GeV.  For model 5$(41/15:1:1)$ the allowed mass range for $m_0$ is $\sim 200-900$
	GeV with $M_3^G$ between $\sim 1300-2000$ GeV for $A_0=0,-1$ TeV, but for $A_0=1$
	TeV, $m_0$ ranges between $\sim 300-800$ GeV with $M_3^G$ between $\sim 1450-2000$
	GeV.}
\label{fig:w1}
\end{figure}

We see that for large $A_0$ the $\stau$ mass becomes very large thereby precluding the
stau-coannihilation channel and as a result the relic density which depends on the stau
coannihilation becomes too large (this holds for  light bino  DM and  applies to model 24
only).  Also very large $\tan\beta$ leads to conflict with the $B_s\to\mu^+\mu^-$
constraint since the SUSY contribution to this process goes as $O(\tan^6\beta)$.

In Table~\ref{tabm2}, model 24 which has a gaugino mass ratio of $-1/2:-3/2:1$ having a
bino LSP at low scale, is compatible with all the low energy constraints considered in
this work.  But it is mainly dependent on the stau coannihilation channel for achieving
the correct relic density which means that one has to choose $m_0$ such that $\stau$ mass
is quasi degenerate with the LSP mass. The sign of $\mu$ is chosen to be negative as that
gives the a positive contribution to $(g-2)$.


Also in Table~\ref{tabm2}, model 20 which has the gaugino mass ratio $5/2:-3/2:1$ having a
higgsino dominated LSP is compatible with all the low energy constraints but only for
$A_0=-1$ TeV.

We show the mass spectrum for wino models in Table~\ref{tab:w}, bino models in
Table~\ref{tab:b} and higgsino models in Table~\ref{tab:h} which satisfy all the low
energy constraints listed in the beginning of the section. These are models $1-5$, $9-11$,
$18-20$, 22 and 24 as given in Tables~\ref{tabm1} and \ref{tabm2}.   The input parameters
for each of the benchmark scenarios are shown in Table~\ref{tab:inpar}. The non-universal
gaugino models 11 and 19 have been examined in ref\cite{11-19}.  For models 11, 19 and 24
the parameter space which satisfies all the constraints is restricted in the neighbourhood
of the values shown in the benchmark table.

\subsubsection*{ Wino DM}

In  models 2$(-3:1:1)$, 3$(-13/5:1:1)$ and 5 $(41/15:1:1)$  the LSP is a wino with mass
1323 GeV, 1073 GeV and 1189 GeV respectively. In all three  models the chargino masses are
almost degenerate with the wino LSP masses due to which the chargino co-annihilation
processes $\tilde \chi_1^0 \tilde \chi_1^+ \to ZW^+, c\bar s, u \bar d$ and  $\tilde
\chi_1^- \tilde \chi_1^+ \to W^-W^+$ make as much contribution to the relic density in
addition as the annihilation channel $\tilde \chi_1^0 \tilde \chi_1^0 \to W^-W^+$. These
models come closest to being probed in the direct detection experiments as discussed in
Section~\ref{Direct-Detection}.  In addition models 1$(-19/5:1:1)$ and 4$(-22/5:1:1)$ also
show a valid parameter space for $M_3=2000-2400$ GeV, this is because of the well known
result that the correct relic density for wino LSP models is achieved at by wino annihilation to W pair
by a t-channel chargino exchange with
$M_{\mbox{LSP}}\sim 2$ TeV \cite{Hisano:2005ec}.

\begin{table}
	\centering
	{\bf Wino Models}\vskip 0.1cm
	\begin{tabular}{|c|c|c|c|c|c|}
        \hline\hline
        Model no. & 1 & 2 & 3 & 4 & 5 \\\hline
	$M_1^G:M_2^G:M_3^G$ & $-\frac{19}{5}:1:1$ & $-3:1:1$ & $-\frac{13}{5}:1:1$ &
	$-\frac{22}{5}:1:1$ & $\frac{41}{15}:1:1$ \\\hline
	$\neut$				&1673	& 1323 	& 1073 	&1688	& 1189  \\           	
        $\nneut$			&2160	& 1852 	& 1514 	&2120	& 1739  \\           	
        $\nnneut$			&2167	& 1861 	& 1606 	&2129	& 1824 \\            	
        $\nnnneut$			&3490	& 2174 	& 1609 	&4071	& 1842 \\            	
        $\charg$			&1673	& 1323 	& 1073 	&1688	& 1189  \\           	
        $\ccharg$			&2168	& 1862 	& 1606 	&2129	& 1829 \\\hline      	
        $M_1$                   	&3538	& 2202 	& 1544 	&4129	& 1776  \\           	
        $M_2$                   	&1632	& 1292 	& 1049 	&1647	& 1160  \\           	
        $M_3$                   	&4144	& 3344 	& 2761 	&4175	& 3028 \\            	
        $\mu$                   	&2149	& 1847 	& 2024 	&2108	& 1818 \\\hline      	
        $\tilde{g}$                  	&4262	& 3431 	& 2835 	&4305	& 3105 \\            	
        $\tilde{\tau}_1$             	&1897	& 1344 	& 1076 	&2080	& 1195  \\           	
        $\tilde{\tau}_2$             	&2835	& 1782 	& 1293 	&3306	& 1482  \\           	
        $\tilde{e}_R,\tilde{\mu}_R$  	&2846	& 1790 	& 1300 	&3318	& 1494  \\           	
        $\tilde{e}_L,\tilde{\mu}_L$  	&1905	& 1349 	& 1081 	&2089	& 1203  \\           	
        $\tilde{t}_1$                	&3330	& 2554 	& 2056 	&3441	& 2218  \\           	
        $\tilde{t}_2$                	&3519	& 2841 	& 2364 	&3585	& 2563 \\            	
        $\tilde{b}_1$                	&3488	& 2822 	& 2344 	&3503	& 2544 \\            	
        $\tilde{b}_2$                	&3717	& 2972 	& 2467 	&3785	& 2699 \\            	
        $\tilde{u}_{R}$          	&4078	& $3160$& $2586$&4255	& $2839$ \\          
        $\tilde{u}_{L}$          	&3834	& $3093$& $2574$&3871	& $2816$ \\\hline    
        $M_h$ (Higgs)		 	& 124	& $123$ & $123$ & 124	& $124$ \\            
        $\Omega h^2$		 	& 0.11	& 0.11  & 0.11  & 0.11	& 0.113\\             
        $a_\mu^{SUSY}~(\times 10^{-10})$& 0.3	& 0.46  & 0.65  & 0.28	& 0.66 \\\hline\hline 
	\end{tabular}
	\caption{The SUSY mass spectrum for a chosen benchmark point as suggested in
	Table~\ref{tab:inpar} for each of the wino models which satisfy all the low energy
	constraints.  In addition we also mention the Higgs mass and the relic density in
	each case.  All masses are in GeV.}
	\label{tab:w}
\end{table}

Of all the wino models only model 8$(77/5:1:1)$ does not have any valid parameter space
for the region that we scan.  Here, for $M_3^G\lesssim1600$ GeV the relic density is under
abundant while for $M_3^G\gtrsim 1600$ there is no EWSB.

\begin{figure}[htp]
	\centering
	\includegraphics[width=0.49\textwidth]{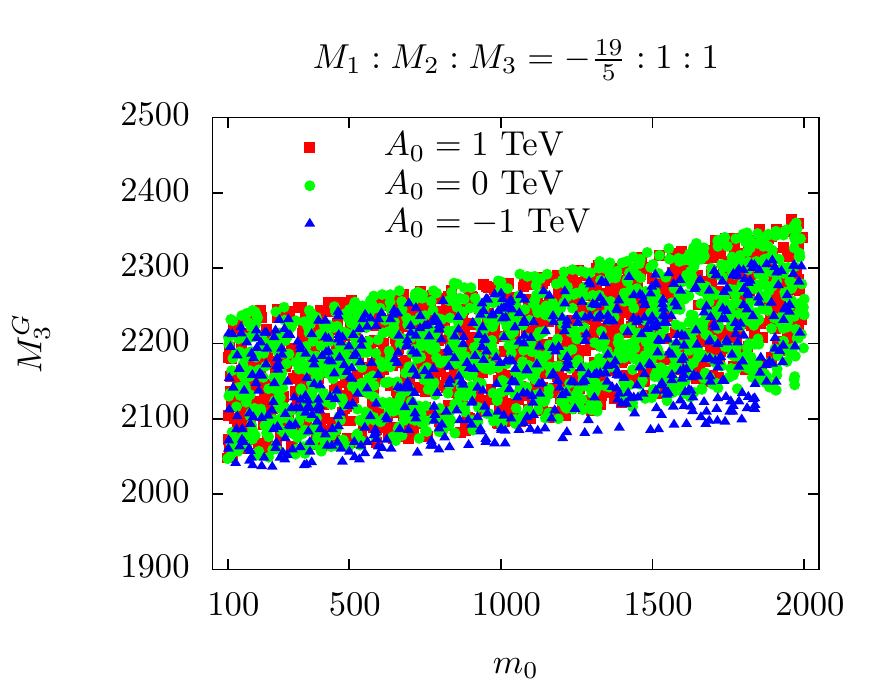}
	\includegraphics[width=0.49\textwidth]{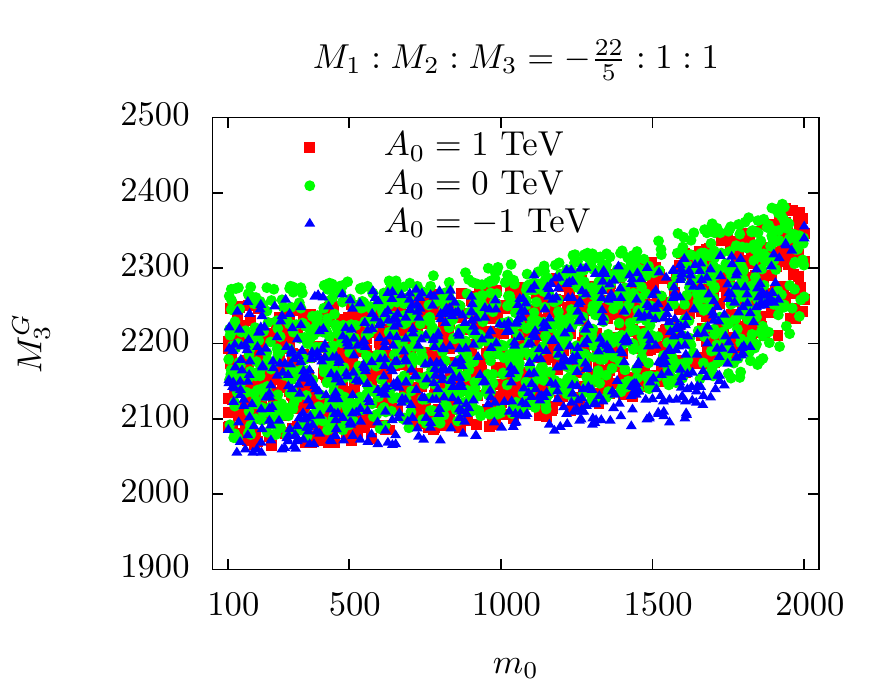}
	\caption{The allowed parameter space for heavy wino models 1$(-19/5:1:1)$ and
	4$(-22/5:1:1)$ shown in the left and right panels respectively.  We extend the
	scan range for $M_3^G$ upto 3 TeV for these two models.  The allowed mass range
	for $M_3^G$ lies between $\sim 2.0-2.4$ TeV while for $m_0$ it covers the entire
	range of our scan from $100-2000$ GeV.}
	\label{fig:w2}
\end{figure}

We have noted that if we allow the larger parameter space for $M_3^G$,  models 1 and
4 allows some parameter space which is consistent with the constraints that we have
imposed in our study. It is interesting to mention that for these models to be compatible
with the correct relic density, $M_3^G$ needs to be more than 2 TeV in both cases, see
Fig.~\ref{fig:w2}. We have not extended $M_3^G$ value beyond 2 TeV for other models as
they already qualify to be allowed models for smaller ranges of parameters.

\subsubsection*{Bino DM}

\begin{figure}[htp]
	\centering
	\includegraphics[width=0.45\textwidth]{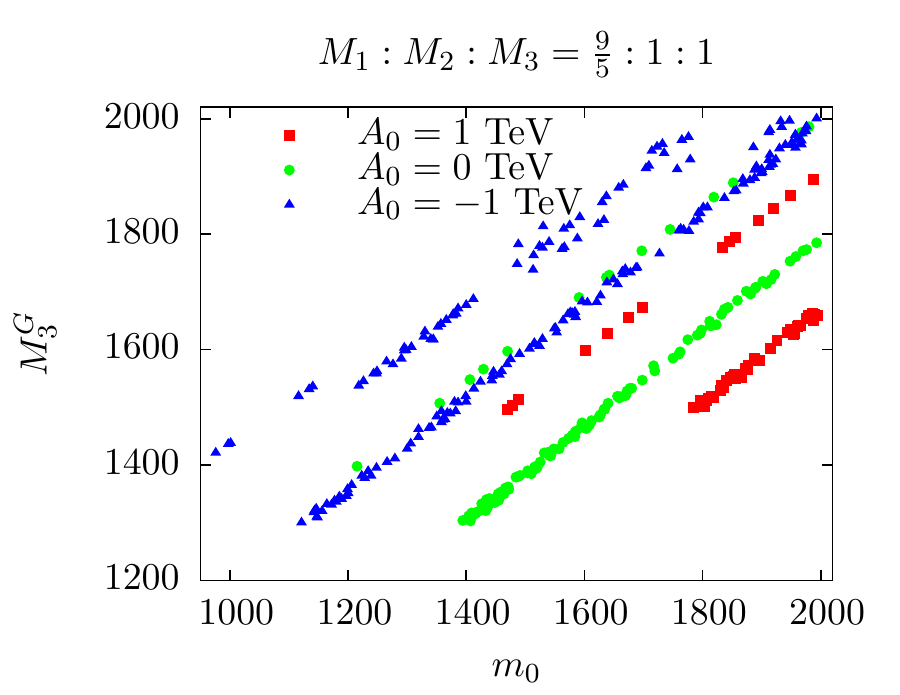}\vskip 0.2cm
	\includegraphics[width=0.45\textwidth]{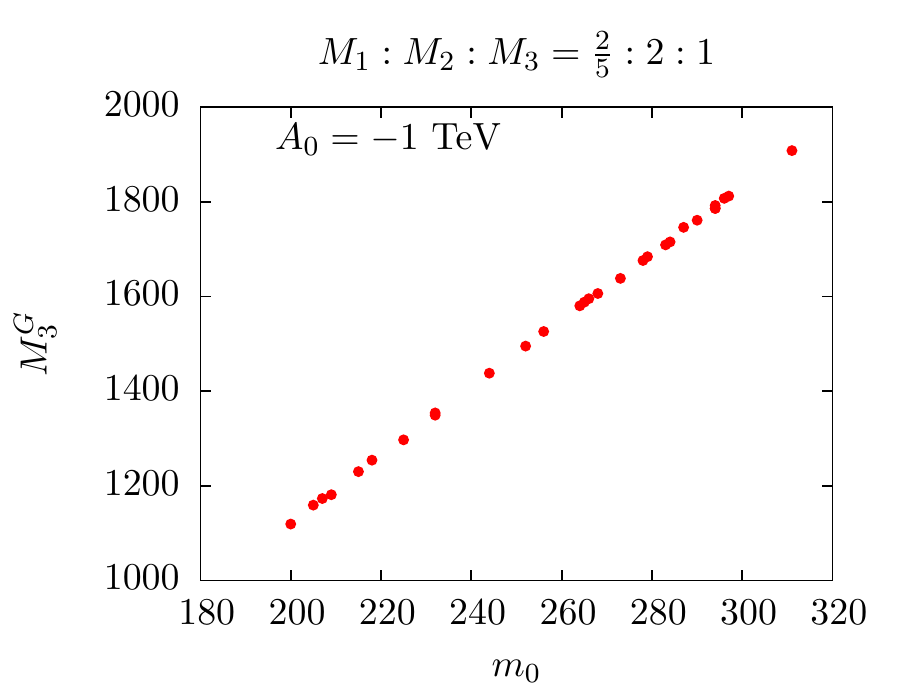}
	\includegraphics[width=0.45\textwidth]{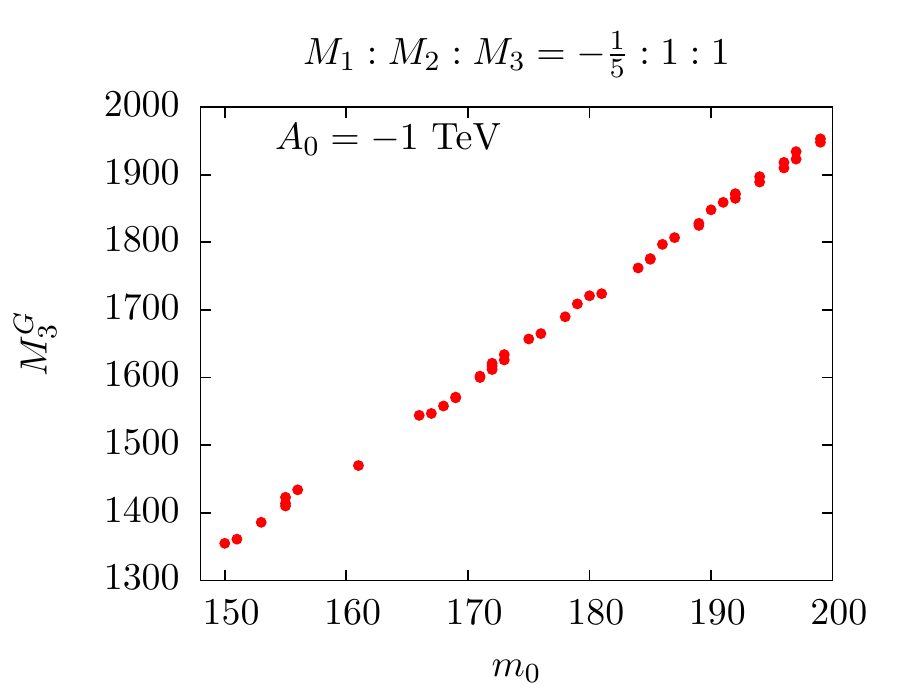}
	\caption{Same as Fig.~\ref{fig:w1} but for heavy bino DM models 10$(9/5:1:1)$,
	18$(2/5:2:2)$ and 22$(-1/5:1:1)$. As before all low energy constraints except muon
	$(g-2)$ are satisfied.  For model 10$(9/5:1:1)$ shown in the top panel, the
	allowed mass range for $m_0$ is $\sim 1200-2000$ GeV with $M_3^G$ between $\sim
	1300-1900$ GeV for $A_0=0$ TeV, while for $A_0=1$ TeV, $m_0$ ranges between $\sim
	1500-2000$ GeV with $M_3^G$ between $\sim 1500-1900$ GeV and finally for $A_0=-1$
	TeV $m_0$ lies between $\sim 1000-2000$ GeV and $M_3^G$ between $\sim 1300-2000$
	GeV.  The models 18$(2/5:2:2)$	and 22$(-1/5:1:1)$ shown in the bottom left and
	right panels respectively, have a small parameter space and are more fine-tuned
	than the other models studied here.  We show the result for $A_0=-1$ TeV.  For
	model 18$(2/5:2:2)$ the allowed mass range for $m_0$ is $\sim 200-320$ GeV and
	$M_3^G$ ranges between $\sim 1100-1900$ GeV.  For model 22$(-1/5:1:1)$ the allowed
	mass range for $m_0$ is $\sim 150-200$ GeV while for $M_3^G$ it is $\sim
	1300-1950$ GeV.}
	\label{fig:b}
\end{figure}

There are three models which have bino LSP as the DM but with very different benchmark
spectrum.  In model 10  $(9/5:1:1)$ the DM is a 934 GeV bino LSP. The chargino mass is
close to the LSP mass and chargino coannihilation processes, $\tilde \chi_1^0 \tilde
\chi_1^+ \to t \bar b$; $\tilde \chi_1^- \tilde \chi_1^+ \to t \bar t, b \bar b$ are
important for relic density.  In addition the Next to Lightest Supersymmetric Particle
(NLSP) mass is close at 970 GeV and the NLSP coannihilation processes, $\tilde \chi_1^0
\tilde \chi_2^0 \to b \bar b$ and $\tilde \chi_2^0 \tilde \chi_2^0 \to b \bar b$ makes a
significant contribution to the DM annihilation.  As a result, the parameters $A_t$
and $A_b$ significantly affect the parameter space for achieving the correct relic
density.  This is seen in the top panel of Fig.~\ref{fig:b} with the parameter space for
different values of $A_0$ being split further apart as compared to Figs.~\ref{fig:w1} and
\ref{fig:w2}.

In model 19 $(-5:3:1)$ the LSP is predominantly bino with  higgsino mixture
($N_{11}=0.826, N_{13}=0.449, N_{14}=0.338$) of mass 159 GeV. The processes $\tilde
\chi_1^0 \tilde \chi_1^0 \to W^+ W^-, Z Z$ contribute to the relic density.

In model 24 $(-1/2:-3/2:1)$  the LSP is a bino of mass 178 GeV and the main annihilation
channel is the stau  coannihilation $\tilde \chi_1^0 \tilde \tau \to A \tau$; $\tilde \tau
\tilde \tau \to \tau \bar \tau, AA$;  $\tilde \chi_1^0 \tilde \tau \to Z \tau$ which are
all an order of magnitude larger than the annihilation channel $\tilde \chi_1^0 \tilde
\chi_1^0 \to \tau \bar \tau$. The stau coannihilation channels are boosted up by taking
the stau mass 184.5 GeV close to the LSP mass.  In addition the models 18 and 22 also show
a very small parameter space in the stau coannihilation region.  These two models in
particular require that the $\stau_1$ mass be taken very close to the LSP mass (within 5
GeV) and in that sense are more fine tuned than the rest of the successful models.

\begin{table}[htp]
	\centering
	{\bf Bino Models}\vskip 0.1cm
	\begin{tabular}{|c|c|c|c|c|c|}\hline\hline
        Model no. & 10 & 18 & 19 & 22 & 24 \\\hline
	$M_1^G:M_2^G:M_3^G$ & $\frac{9}{5}:1:1$ & $\frac{2}{5}:2:1$ & $-5:3:1$ & 
	$-\frac{1}{5}:1:1$ & $-\frac{1}{2}:-\frac{3}{2}:1$ \\\hline
	$\neut$				 & 934.3 &188.6	 & 159.2 &131.2	 & 177.6 \\
        $\nneut$			 & 970.4 &1252	 & 202.6 &1103	 & 976.4 \\
        $\nnneut$			 & 1551  &1259	 & 219.6 &1696	 & 1523 \\
        $\nnnneut$			 & 1558  &1828	 & 4219  &1699	 & 1528 \\
        $\charg$			 & 970.1 &1252	 & 1999  &1103	 & 976.4 \\
        $\ccharg$			 & 1557  &1828	 & 4219  &1699	 & 1528 \\\hline
        $M_1$                   	 & 943.6 &190.1	 & 174.4 &133.2	 & 177.9 \\
        $M_2$                   	 & 943.1 &1803	 & 4194  &1076	 & 981.4 \\
        $M_3$                   	 & 2497  &2344	 & 3494  &2824	 & 1771 \\
        $\mu$                   	 & 1545  &1251	 & 1943  &1691	 & 1521 \\\hline
        $\tilde{g}$                  	 & 2596  &2420	 & 3644  &2883	 & 1805 \\
        $\tilde{\tau}_1$             	 & 1253  &195.3	 & 740.6 &139.1	 & 184.5 \\
        $\tilde{\tau}_2$             	 & 1301  &1419	 & 3310  &857.6	 & 861.2 \\
        $\tilde{e}_R,\tilde{\mu}_R$  	 & 1271  &259.4	 & 797.3 &184.6	 & 528.0 \\
        $\tilde{e}_L,\tilde{\mu}_L$  	 & 1303  &1424	 & 3316  &861 	 & 926.6 \\
        $\tilde{t}_1$                	 & 1847  &1391	 & 1559  &1959	 & 775.7 \\
        $\tilde{t}_2$                	 & 2264  &2258	 & 4039  &2398	 & 1440 \\
        $\tilde{b}_1$                	 & 2250  &2049	 & 3089  &2386	 & 1404 \\
        $\tilde{b}_2$                	 & 2412  &2249	 & 4033  &2465	 & 1473 \\
        $\tilde{u}_{R}$          	 & $2467$&2069	 & $3125$&2480	 & $1632$ \\
        $\tilde{u}_{L}$          	 & $2515$&2482	 & $4439$&2606	 & $1800$ \\\hline
        $M_h$ (Higgs)		 	 & $123$ & 123	 & $125$ & 122	 & $124$ \\
        $\Omega h^2$		 	 & 0.12  & 0.11	 & 0.12  & 0.11	 & 0.12 \\
        $a_\mu^{SUSY}~(\times 10^{-10})$ & 0.79  & 0.16	 & 0.28  & 1.0	 & 2.65 \\\hline\hline
	\end{tabular}
	\caption{The SUSY mass spectrum for a chosen benchmark point as suggested in
	Table~\ref{tab:inpar} for each of the bino models which satisfy all the low energy
	constraints.  In addition we also mention the Higgs mass and the relic density in
	each case.  All masses are in GeV.}
	\label{tab:b}
\end{table}

\begin{figure}[htp]
\centering
	\includegraphics[width=0.45\textwidth]{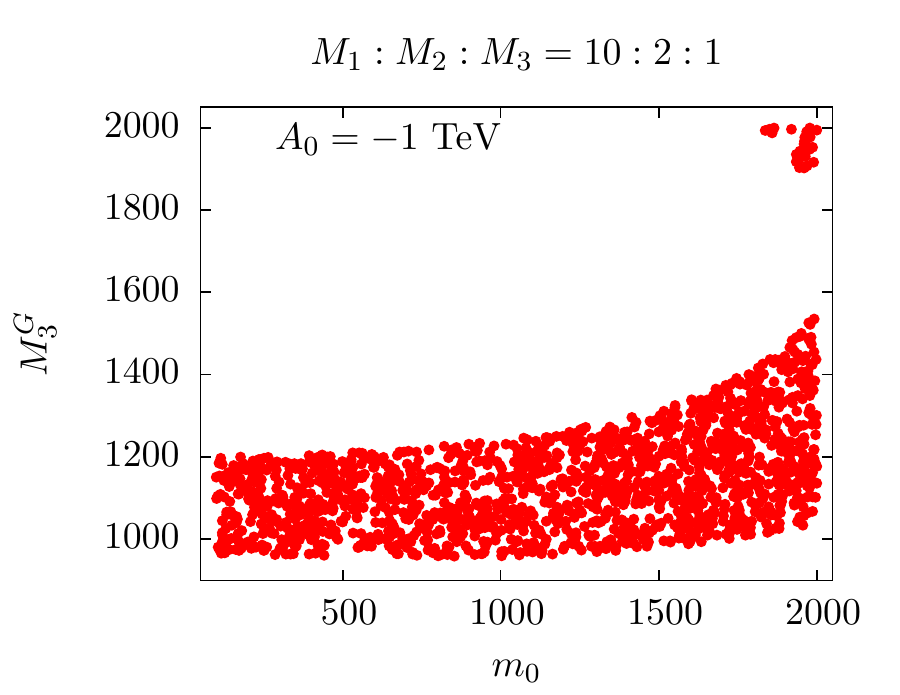}
	\includegraphics[width=0.45\textwidth]{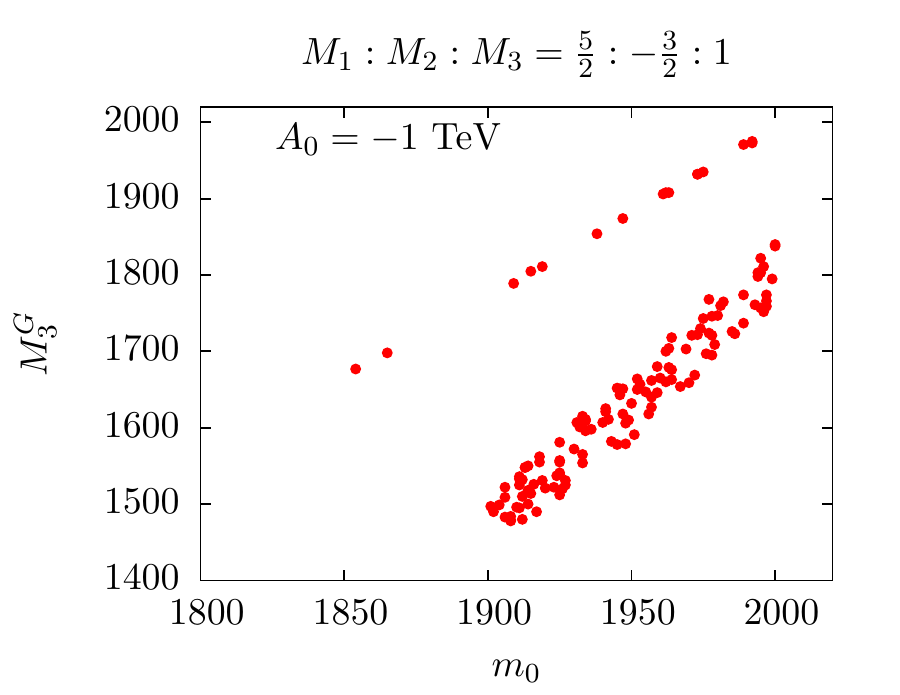}
\caption{The allowed parameter space satisfying the low energy constraints except muon
	$(g-2)$ for heavy higgsino DM models 9$(10:2:1)$ and model 20$(5/2:-3/2:1)$. All
	parameters are chosen as in Fig.~\ref{fig:w1} except $A_0=-1$ TeV.  For model
	9$(10:2:1)$ the allowed	mass range for $m_0$ spans the entire range of scan from
	$100-2000$ GeV with $M_3^G$ between $\sim 950-1550$ GeV.  For model
	20$(5/2:-3/2:1)$ the allowed mass range for $m_0$ is $\sim 1850-2000$ GeV with
	$M_3^G$ between $\sim 1400-2000$ GeV.  These models do not work for $A_0=0,1$
	TeV.}
\label{fig:h}
\end{figure}

The bino models which do not work in our parameter scan are models 14, 15, 16, 17, 18, 21,
22, 23 and 25 with their ratios as given in Table 2.  For these models either the relic
density is over abundant or $\stau_1$ becomes the LSP or the model is unphysical
(tachyonic modes).  For model 14, when $m_0\leqslant200$ GeV $\stau_1$ is the LSP, and
when $m_0>200$ GeV the relic density is over abundant with stau coannihilation dominating
in the lower $m_0$ range while for $m_0>500$ GeV the dominant contribution to relic
density coming from leptonic channel which is suppressed.  In model 15 the correct relic
density is achieved through stau coannihilation for $m_0\lesssim 200$ GeV, but the Higgs
mass is lighter than the acceptable limit of 122 GeV.  Whereas for $m_0\gtrsim 200$ GeV
the Higgs mass is in the acceptable range for most of the parameter space but the relic
density becomes overabundant with annihilation to leptons dominating the relic density
contribution.  In addition for $M_3^G>1$ TeV $\stau_1$ becomes the LSP.  In model 16, for
$m_0<200$ GeV the parameter space is unphysical, and for $m_0\geqslant 200$ the relic
density is over abundant with the dominant annihilation channels into  $\tau\bar\tau$ and
$b\bar b$.  Model 17 is similar to model 15, however in case of model 17 the $\stau_1$
mass is below LEP limit for $m_0<200$ GeV.  Model 21 is similar to model 18, but is ruled
out because of the higgs mass constraint.  For model 22, at low $m_0$ values below $400$
GeV the LSP is $\stau_1$.  At higher values of $m_0$ the bino LSP which gives overabundant
relic density crosses over to higgsino dominated LSP as $M_3$ increases.  For the region
with higgsino dominated LSP the relic density is again overabundant with the main
contribution to relic density coming from coannihilation channel.  Model 25 behaves
similar to model 22, however for low $m_0$ values below 300 GeV the correct relic density
is achieved through stau coannihilation however the higgs mass constraint is not
satisfied.  While for higher values of $m_0$ beyond $300-400$ GeV, the higgs mass
constraint does get satisfied but the relic density remains overabundant even in
stau-coannihilation region of the parameter space.

\subsubsection*{ Higgsino DM}

In model 9 $(10:2:1)$ the LSP is a higgsino and the relic density is via the chargino
coannihilation processes $\tilde \chi_1^0 \tilde \chi_1^+ \to u \bar d, c \bar s$.
The NLSP mass is close to the LSP mass and the NLSP coannihilation $\tilde
\chi_2^0 \tilde \chi_1^+ \to u \bar d, c \bar s$ also contributes to the relic density.

In model 11 $(-1/5:3:1)$ the LSP is a higgsino with mass  1015 GeV and the relic density
is via the same chargino coannihilation processes as in model 9 including the NLSP
coannihilation contribution.

\begin{table}
	\centering
	{\bf Higgsino Models}\vskip 0.1cm
	\begin{tabular}{|c|c|c|c|}
	\hline\hline
        Model no. & 9 & 11 & 20 \\\hline
	$M_1^G:M_2^G:M_3^G$ & $10:2:1$ & $-\frac{1}{5}:3:1$ & $\frac{5}{2}:-\frac{3}{2}:1$
	\\\hline
	$\neut$				&1006	& 1015     & 1507      \\
        $\nneut$			&1013	& 1016     & 1510      \\
        $\nnneut$			&1584	& 3791     & 1958     \\
        $\nnnneut$			&4258	& 4093     & 2230     \\
        $\charg$			&1007	& 1015     & 1507      \\
        $\ccharg$			&1584	& 4093     & 2230     \\\hline
        $M_1$                   	&4294	& 3797     & 1969      \\
        $M_2$                   	&1549	& 4051     & 2175      \\
        $M_3$                   	&2023	& 3361 	   & 3570     \\
        $\mu$                   	&1002	& 1000 	   & 1495     \\\hline
        $\tilde{g}$                  	&2164	& 3585 	   & 3772     \\
        $\tilde{\tau}_1$             	&2138	& 3181 	   & 2455      \\
        $\tilde{\tau}_2$             	&3537	& 3779 	   & 2619      \\
        $\tilde{e}_R,\tilde{\mu}_R$  	&3554	& 3620 	   & 2473      \\
        $\tilde{e}_L,\tilde{\mu}_L$  	&2152	& 3181 	   & 2628      \\
        $\tilde{t}_1$                	&1767	& 2309 	   & 2910      \\
        $\tilde{t}_2$                	&2254	& 3716 	   & 3625     \\
        $\tilde{b}_1$                	&1782	& 2812 	   & 3617     \\
        $\tilde{b}_2$                	&2125	& 3726 	   & 3656     \\
        $\tilde{u}_{R}$          	&2945	& $4051$   & $3785$    \\
        $\tilde{u}_{L}$          	&2226	& $4662$   & $3988$    \\\hline
        $M_h$ (Higgs)		 	& 124	& $127$    & $122$     \\
        $\Omega h^2$		 	& 0.11	& 0.12     & 0.11     \\
        $a_\mu^{SUSY}~(\times 10^{-10})$& 0.44	& 0.47     & 0.24     \\\hline\hline
	\end{tabular}
	\caption{The SUSY mass spectrum for a chosen benchmark point as suggested in
	Table~\ref{tab:inpar} for each of the higgsino models which satisfy all the low
	energy constraints.  In addition we also mention the Higgs mass and the relic
	density in each case.  All masses are in GeV.}
	\label{tab:h}
\end{table}

In model 20 $(5/2: -3/2:1)$ the LSP is a higgsino of mass 1507 GeV and the contributions
to the relic density are due to the chargino  coannihilation $\tilde \chi_1^0 \tilde
\chi_1^+ \to t \bar b$; $\tilde \chi_1^- {\tilde \chi_1}^+ \to t \bar t, ~b \bar b $ in
addition to the main annihilation channel $\tilde \chi_1^0 \tilde \chi_1^0 \to b \bar b,~
t \bar t$. The NLSP mass is close to the LSP mass and the NLSP coannihilation $\tilde
\chi_2^0 \tilde \chi_1^+ \to t \bar b$ also contributes to the relic density. This model
gives the correct relic density for $A_0 \sim -1$ TeV.

The failed higgsino models are models 6($122/5:1:1$), 7($-101/10:-3/2:1$), 12($1:35/9:1$)
and 13($1:-5:1$).  All of these models fail because the spectrum is unphysical or the
higgs sector is unstable.  In model 6 for $m_0\leq 1200$ GeV the spectrum contains
tachyonic modes, while for $m_0\geq 1200$ GeV there is no EWSB and as $M_3$ increases one
again encounters tachyonic modes in the spectrum. In model 7 the relic density is under
abundant for $M_3^G<1.3$ TeV while for higher values of $M_3^G$ there is no EWSB.  Model
12 behaves very similar to model 6 and so fails for the same reasons.  For model 13, there
is no EWSB below a certain value of $M_3$ for a given $m_0$, and this value increases with
$m_0$.  Above this value of $M_3$ some of the scalar modes are tachyonic .

\subsection{Direct detection constraints}\label{Direct-Detection}

 \begin{figure}[htp]
\centering
	\includegraphics[width=0.45\textwidth]{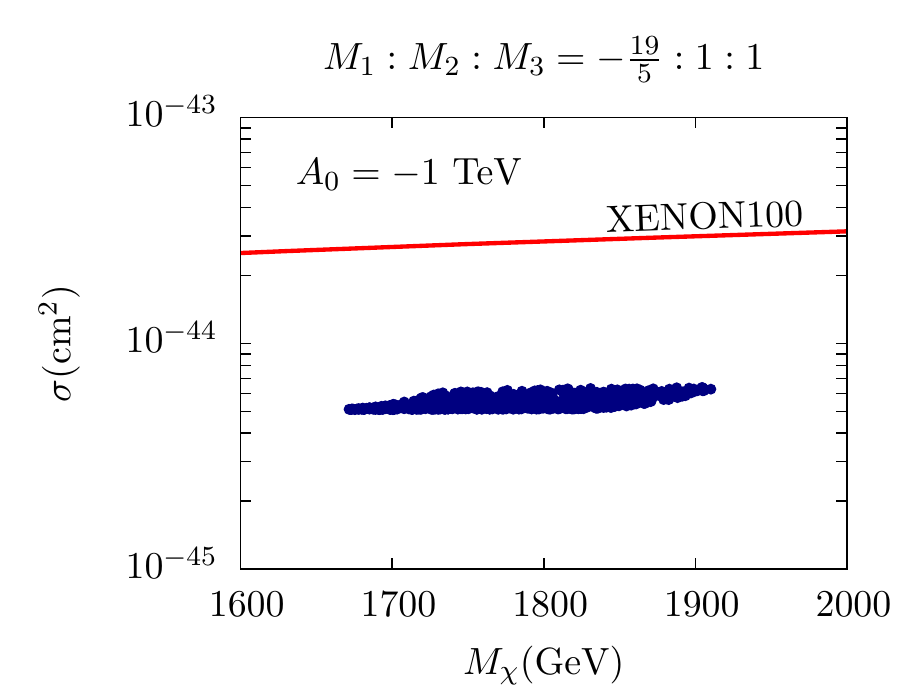}
	\includegraphics[width=0.45\textwidth]{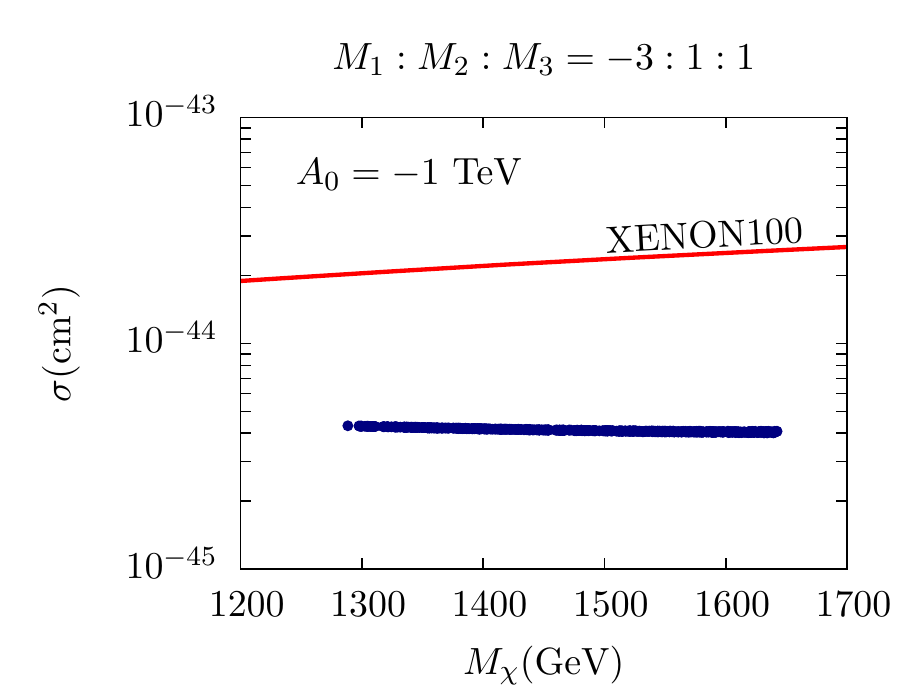}\vskip 0.2cm
	\includegraphics[width=0.45\textwidth]{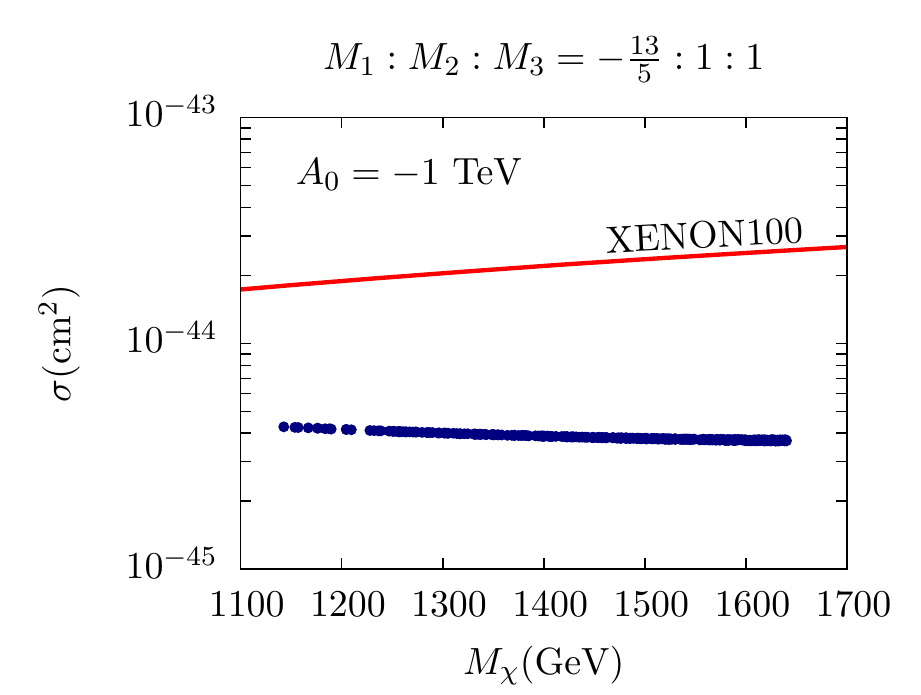}
	\includegraphics[width=0.45\textwidth]{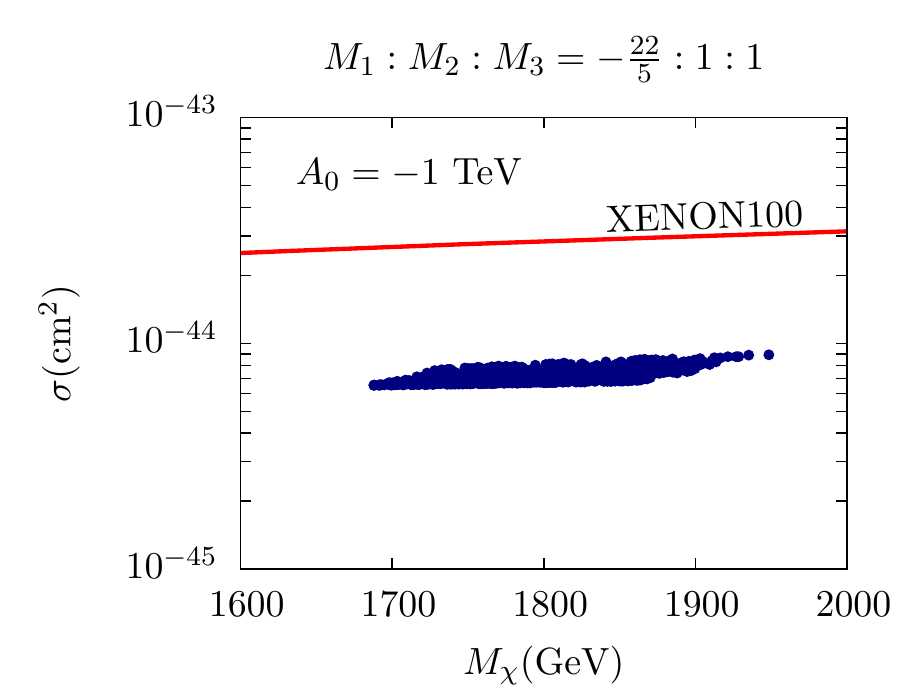}\vskip 0.2cm
	\includegraphics[width=0.45\textwidth]{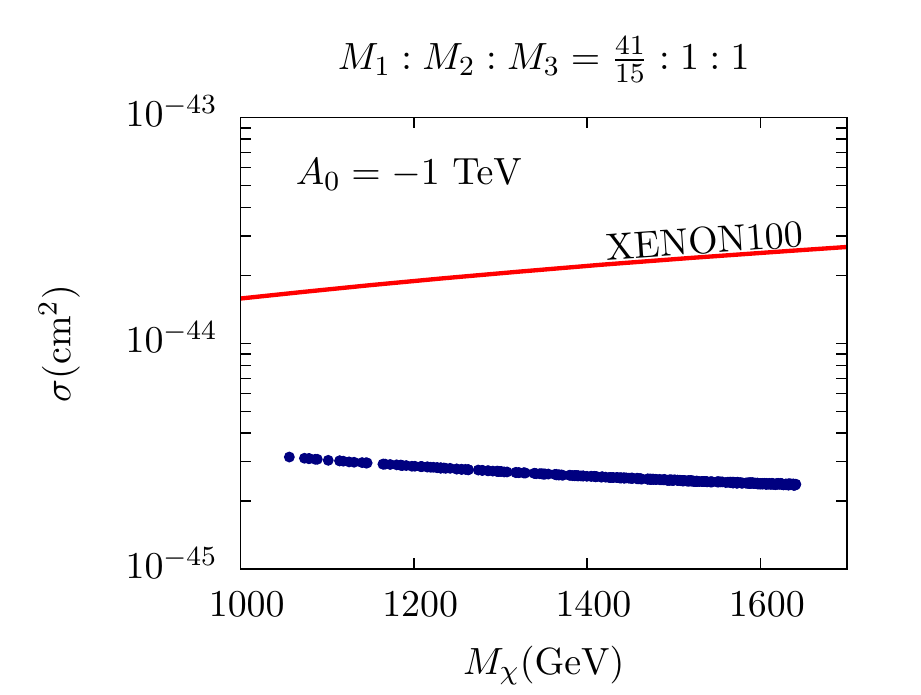}
\caption{The direct detection spin independent proton-DM scattering cross section plotted
	with the constraint from XENON100 \cite{xenon100}.  These plots show selected
	points for the heavy wino models satisfying all the low energy constraints
	considered here, except for muon $(g-2)$.  These heavy wino models satisfy the
	XENON100 constraint.}
\label{dd1}
\end{figure}

\begin{figure}[htp]
\centering
	\includegraphics[width=0.45\textwidth]{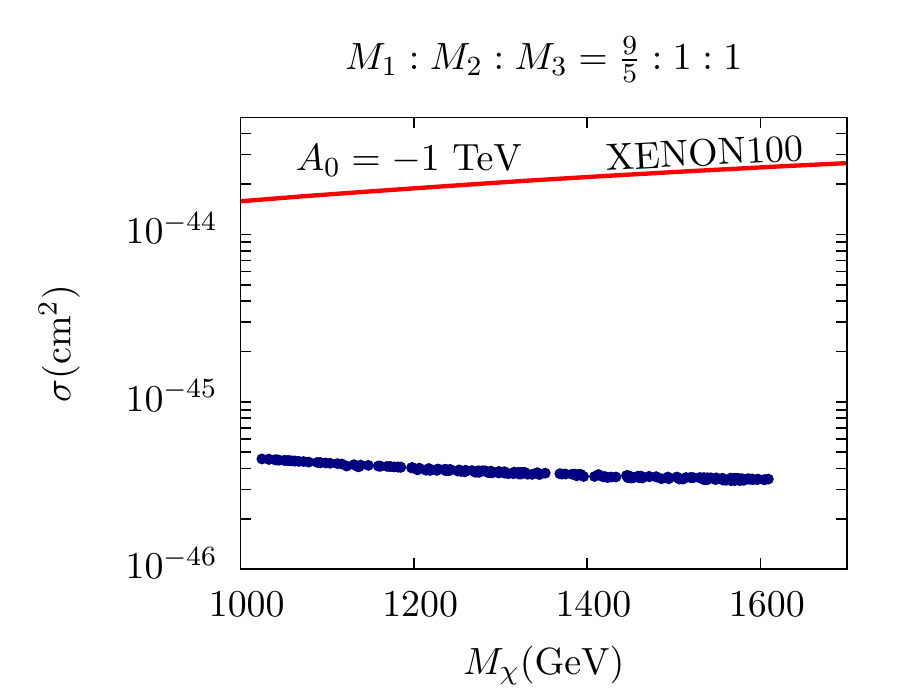}
	\includegraphics[width=0.45\textwidth]{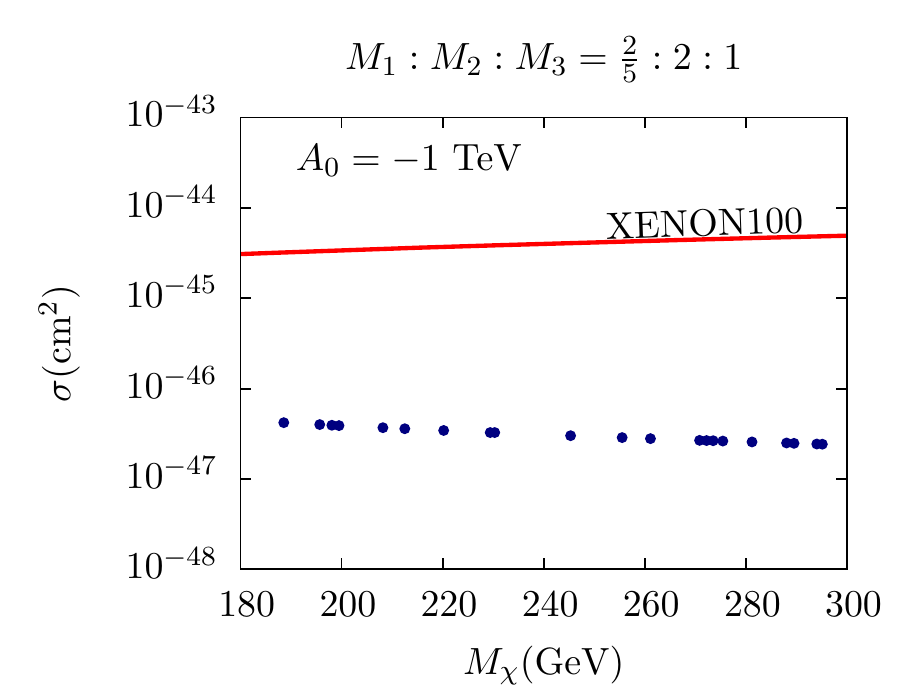}\vskip 0.2cm
	\includegraphics[width=0.45\textwidth]{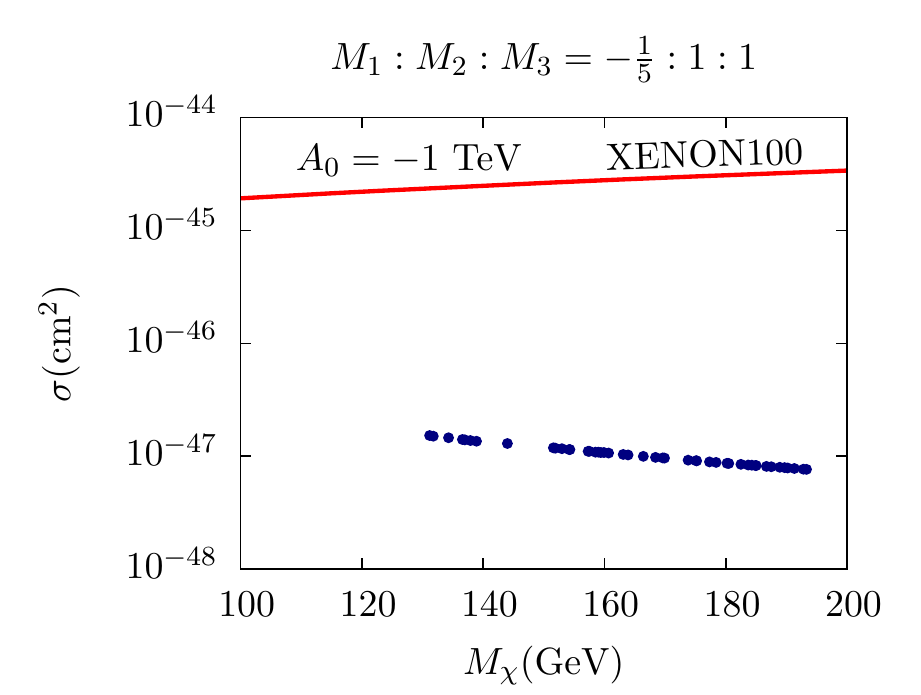}
\caption{The direct detection spin independent proton-DM scattering cross section plotted
	with the constraint from XENON100 \cite{xenon100}.  These plots show selected
	points for bino models satisfying all the low energy constraints considered here,
	except for muon $(g-2)$.  These bino models satisfy the XENON100 constraint.}
\label{dd2}
\end{figure}

\begin{figure}[t]
\centering
	\includegraphics[width=0.45\textwidth]{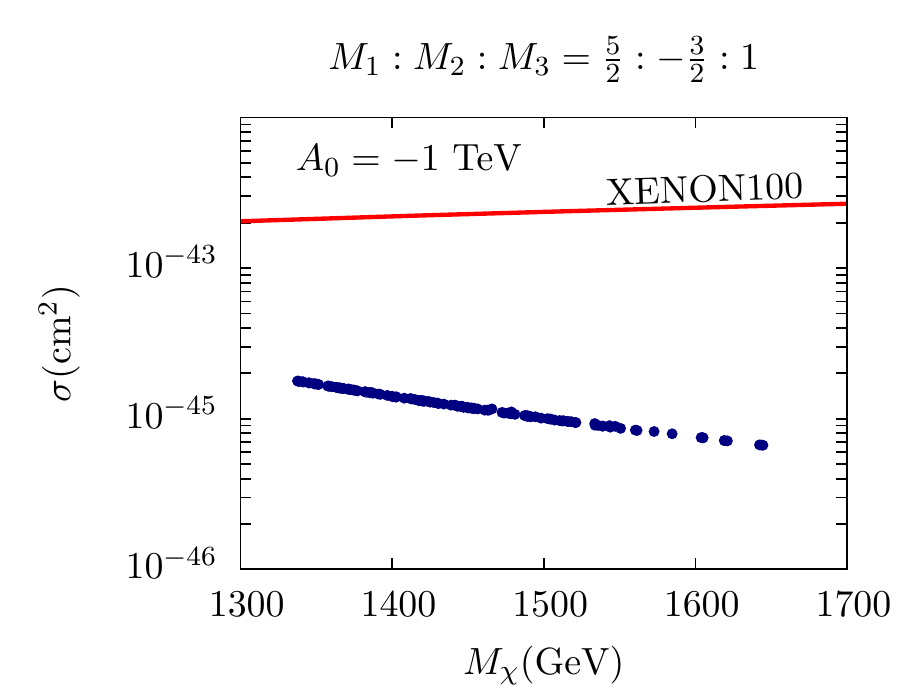}
	\includegraphics[width=0.45\textwidth]{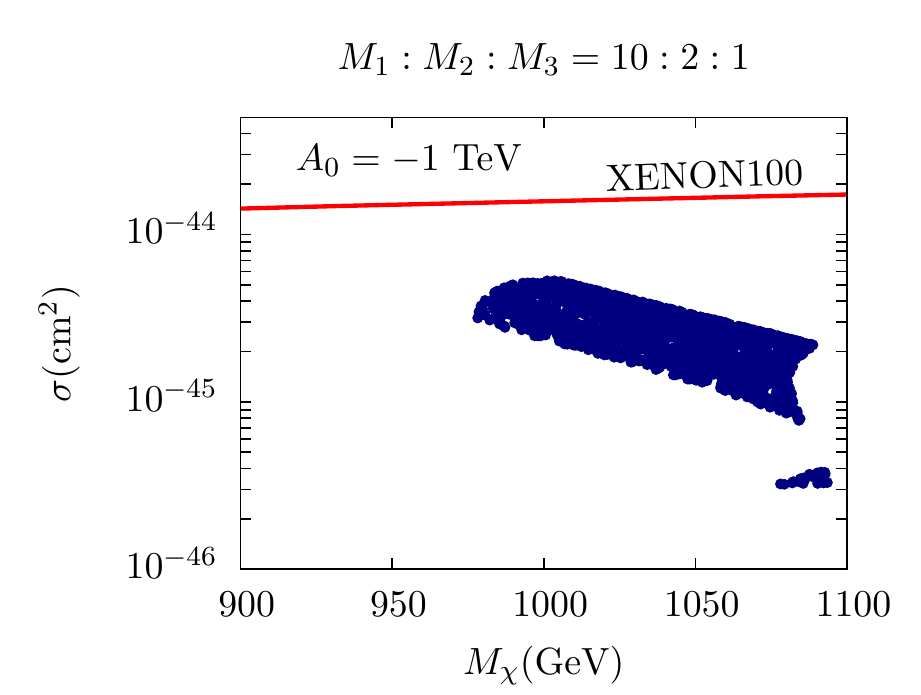}
\caption{The direct detection spin independent proton-DM scattering cross section plotted
	with the constraint from XENON100 \cite{xenon100}.  These plots show selected
	points for the heavy higgsino models satisfying all the low energy constraints
	considered here, except for muon $(g-2)$.  These heavy higgsino models satisfies
	the XENON100 constraint.}
\label{dd3}
\end{figure}

The elastic scattering of neutralinos with nucleons which results in spin-independent
cross section is by Higgs exchange. The Higgs coupling to the lightest neutralino depends
upon the product of the higgsino and the gaugino fraction of the neutralino. Pure bino DM
therefore easily evade the direct detection limits from XENON100 \cite{xenon100}.  In
model 24 ( $5/2:-3/2:1$) with a 176 GeV bino DM evades the XENON100 bound but may be
probed in Xenon 1000 as shown in Fig~\ref{dd3}. While model 10 ($9/5:1:1$) which gives a
$\sim 1$ TeV bino DM also easily evades the XENON100 bound as shown in Fig~\ref{dd2}. In
model 19 $(-5:3:1)$ where the 159 GeV LSP is predominantly bino  with a higgsino mixture
($N_{11}=0.826, N_{13}=0.449, N_{14}=0.338$) has a SI cross section $\sim 1.01 \times
10^{-8} {\rm pb}$ and is incompatible with the XENON100 exclusion limits.

The Spin Independent (SI) cross section for model 20 ($5/2:-3/2:1$) which is a 1.5 TeV
higgsino DM also evades the XENON100 bound easily as shown in Fig~\ref{dd2} as the gaugino
fraction is small. Similarly model 11$(-1/5:3:1)$ with a 1 TeV wino DM has a SI cross
section $\sim 7 \times 10^{-11} {\rm pb}$ and evades the XENON100 bound.

The three wino dark matter models 2 ($-3:1:1$), 3 ($-13/5:1:1$) and 5 ($41/15:1:1$) with a
small mixing of higgsino have larger SI  cross sections as shown in Fig~\ref{dd1}.  These
wino DM models may be within the reach of XENON1T \cite{Xeno1T} and Super-CDMS
\cite{superCDMS} experiments.

\section{Muon $(g-2)$}
It has long been recognised that to explain the discrepancy between
experiment and SM prediction for muon anomalous magnetic moment from
a SUSY contribution would require a light mass spectrum on the
gauginos and the sleptons \cite{g-2Martin,g-2Stockinger} which would
put a severe restriction on the SUSY models.

The SUSY contribution to muon $(g-2)$ for light binos is through the bino-smuon loop
\cite{Endo:2013bba,Cho:2011rk} so the largest $a_\mu^{SUSY}=2.65 \times 10^{-10}$
\cite{mug-2,g-2} comes from model 24 which has the lightest LSP (177 GeV bino) and slepton
spectrum. In model 24 $(M_1^G:M_2^G:M_3^G=-1/2:-3/2:1)$ it would have been easy to adjust
the smuon mass (through $m_0^G$) and the bino mass through $M_3^G$ (as $M_1^G$ is related
to $M_3^G$) to get a much larger contribution to muon $(g-2)$. However the relic density
of bino DM in model 24 depends on the stau coannihilation which has to be close to the
bino DM mass of 177 GeV which is again determined by the universal scalar mass $m_0$. So
demanding the correct relic density results in a less than optimum contribution to the
muon $(g-2)$ in this model. In Table~\ref{tabm2} one can note that the gaugino mass ratio
referred to here as model 24, can arise from three possible breaking patterns of $SO(10)$,
each of them through a different intermediate symmetry group. It will be interesting to
see if we distinguish intermediate scale separately than the unification scale then muon
$(g-2)$ is further improved or not. We have kept this issue for our further publication.
The gaugino mass ratio of model 24 has been studied in ref.~\cite{shafi} in the context of
Yukawa unification in $SO(10)$, but in the benchmark models examined in \cite{shafi} the
SUSY contribution to muon $g-2$ is an order of magnitude  smaller than the benchmark
parameters for model 24 shown in Table~\ref{tab:b}.

   In this paper we have chosen a single non-singlet scalar for giving masses to the
   gauginos.  By choosing a the gaugino masses to arise from more than one scalar
   representation like 1+24, 1+75 and 1+200 of $SU(5)$ \cite{King:2007vh, debottam,
   rao-roy1} it is possible to explain muon $(g-2)$ from SUSY contributions along with the
   Planck-WMAP relic density \cite{rao-roy2}. It has been noted \cite{Akula:2013ioa} that
   in a mSUGRA model  the gaugino mass ratio $M_1:M_2:M_3=1:1:10$ at the GUT scale gives
   the required muon $(g-2)$, but in this paper we see that this gaugino ratio does not
   arise from any of the GUT breaking patterns if one considers one non-singlet Higgs
   representation for generating the gaugino masses.

If one were to have non-universal scalar masses
\cite{Miller:2013jra, Badziak:2013eda} it may be possible to adjust
the stau mass to control the relic  relic density and the smuon mass
to fit muon $(g-2)$ using a single scalar representation for getting
non-universal gaugino masses.

\section{Conclusions}

In this paper we have exhaustively analysed all possible non-universal gaugino mass models
that arise from $SU(5),SO(10),E(6)$ SUSY GUT models. The underlying assumption is that the
full gauge symmetry is broken to the SM symmetry group at the GUT scale itself, i.e., the
intermediate scales are same as the GUT scale.  We have considered all these models in its
minimal versions, i.e., we have not probed the effect of the presence of multiple
non-singlet scalars. If one considers that the contribution to the effective gaugino mass
ratios are outcome of the contributions from more than one scalar field with the
introduction of one or more  free parameters, the the unique group theoretic
characteristics of the models are lost.  Thus we restrict ourselves to the minimal
versions (from the point of number of free parameters) of the non-universal gaugino
models.  We have shown different models predict different kind of LSP compositions. Thus
the contributions to the relic density from such models are discriminated. We have
performed a comparative study among such models using the collider constraints, lightest
Higgs mass and the  relic density. We also emphasise the importance of muon $(g-2)$ and
briefly argue why model 24 ($M_1^G : M_2^G : M_3^G = -1/2 : -3/2 : 1$) is the best
candidate among other models in the context of muon $(g-2)$ contribution. We also check
the status of bino-, wino-, and higgisno- dominated models in the context of Direct
detection constraints. The model 19 $(-5:3:1)$ is ruled out by XENON100 \cite{xenon100}.
The three models 2$(-3:1:1)$, 3$(-13/5:1:1)$ and 5$(41/5:1:1)$ where the dark matter is a
TeV scale wino can be probed in upcoming direct detection experiments like XENON1T
\cite{Xeno1T} and Super-CDMS \cite{superCDMS}.

Finally we would like to comment on the impact of the insertions of the intermediate
scales.  In supersymmetric grand unified theories in case of one step breaking the usual
trend of the intermediate scale is to lie around the unification scale, see
\cite{Chakrabortty:2009xm}. Thus we expect that the ratios at the GUT scale will not
change visibly by the new set of RGEs from intermediate scale to the unification scale.
But in case of two step symmetry breaking the second intermediate scale can as low as 100
TeV \cite{Chakrabortty:2009xm} within a proper unification frame work. If the second
intermediate scale is low enough then a new set of RGEs will change the gaugino mass
ratios at the GUT scale widely. We are looking into this issue in detail and postpone and
will present the results in a future publication.

\section*{Acknowledgement}

Work of JC is supported by Department of Science \& Technology, Government of INDIA under
the Grant Agreement number IFA12-PH-34 (INSPIRE Faculty Award).


\end{document}